# Governance for Security, Risks, Competition and Cooperation – Mapping the knowledge

Julián David Cortés-Sánchez[1,2]

Diego Alonzo García-Bonilla[3]

Edgar Guillermo Rodriguez Guevara[3]

Diana Lorena Pineda Ospina[3]

## 1   Introduction

Governance is associated with multiple ways of governing (i.e., directing, guiding, or regulating) at the national, regional, organizational, or individual levels, among several diverse and divergent definitions (Robichau, 2011). Further factors such as security and risks, competition, and cooperation, certainly fit within the governance framework-reach. For instance, research in the lines above has focused on the variation of risk-taking actions of banks with the power of shareholders within the corporate governance (Laeven & Levine, 2009), the development of 'security governance' concept to understanding the European security coordination, management, and regulation, policy framework (Webber et al., 2004), and the role of governance and competition in firms' innovation (Ayyagari et al., 2011). Such exploration leads toward a bountiful research agenda on governance and: security or risks or competition or cooperation, nurtured by multiple problems and multi-inter-transdisciplinary approaches.

Nevertheless, several questions arise when research documents on a given topic exceed a dozen and account for hundreds or even thousands. Those questions might be, among others, what are the emergent and consolidated research topics? What topics are central or already peripherical? To what research fields are those topics related? Is there an interconnectedness between research fields? Such questions are a matter of study of bibliometrics. Bibliometrics is a field aimed to understand the volume, impact, and structure of scientific and technological literature from a quantitative perspective (Pritchard, 1969). Such insights enable to identify drivers for science, technology, and innovation development, and the design and evaluation of national, regional, and organizational research policies (Fortunato et al., 2018).

Bibliometric perspectives on governance and risks or competition or cooperation, highlight multiple approaches to the definition of governance (Boyer-Villemaire et al., 2014; Eulerich et al., 2013; Lebret, 2015; Lidskog & Lidskog, 2017). Da Silva and colleagues (2011) raised the scope of the conceptualization of how governance goes beyond the board of directors, expanding the role of the organization in society. The conceptualization of governance, following Nielsen and Faber's (2019) statements, has three approaches toward risk, sustainability, and resilience. Despite these findings, they lack an integrative framework for governance and its interconnection with risks, competition, and cooperation.

With all the above, our *ad-hoc* contribution and aim are to generate a map of the knowledge based on the research on topics related to governance and security, risks, competition and cooperation for the FDDI (Fudan Development Institute) proceedings publishing project: *'Reflections on Governance: Security and Risks, Competition and Cooperation.'* That mapping exercise would enable a broader audience to delve into the current state, and interdisciplinary pathways of the research published worldwide for addressing complex problems of governance. Following this introduction, the second section presents the bibliometric methods used and the results' interpretation. The third section presents the results, followed by the fourth and fifth sections of discussion and conclusion, respectively.

[1]Principal professor, School of Management and Business, Universidad del Rosario, Colombia.
[2]Invited researcher, Fudan Development Institute, Fudan University, China.
[3]Professor, Faculty of Administration Sciences, Universidad del Valle, Colombia.

## 2 Methods

*2.1 Data*

We gathered data from Elsevier's Scopus. Scopus contains more than 75 million items (i.e., articles, proceedings, and books) authored by 16 million authors affiliated with more than 70,000 institutions (Scopus, 2019). We search for research articles published between 1998-2018 with titles with the keywords' *governance'* and *'security'* or *'risk'* or *'competition'* or *'cooperation.'* Articles' title was chosen over authors' keywords to ensure the importance of the topics in each document (Nakamura et al., 2019). More than 2,600 authors published 1,315 articles that met the search criteria.

*2.2 Methods*

We applied several bibliometric techniques for knowledge mapping, namely: superposition map, thematic evolution, and bibliographic coupling network (BCN). Here we present the methods aims briefly:

- The superposition map calculates the stability measures of a given field by identifying the growth associated with the number of keywords

- The thematic evolution investigates the development of analytical categories

- The BCN is used for both predict and describes emergent research topics since it pictures the current state of research in a field (Boyack & Klavans, 2010; Zhao & Strotmann, 2008). We also conducted a network analysis for the BCN (see supplementary material, part 1: equations and definitions, and part 2: cluster's top-five nodes according to betweenness centrality). Three properties were discussed for the BCN, balancing both exploratory and explanatory features: networks' density and clusters' composition, and nodes' betweenness.

    A network is composed of nodes (i.e., spheres) and edges (i.e., links). In the BCN, the nodes are articles, and two articles are connected if they use similar bodies of knowledge (*e.g.,* articles, books, or proceedings) (Figure 4). A network's density is the proportion of links in a network relative to the total number of links possible. A density of 1 means the whole network is connected (Scott, 1988). A cluster's composition is the percentage of nodes over the total of nodes in a network that composed a given cluster (Bastian et al., 2009). If a cluster shows 25%, it means that it is composed of one-quarter of all the nodes. The betweenness measures how important a node is to the shortest paths through the network (Opsahl et al., 2010). It captures a node's capacity in allowing information to pass from one part of the network (i.e., cluster) to the other (Golbeck, 2015). Simply put, an article with high betweenness is likely to be aware of what is being researched (i.e., *state of the art*) in multiple research circles. Sort of the more 'multi-inter-transdisciplinary' or 'usable' actor of the cluster/network.

We provide the following link for complete access to the network analysis: [link pending to be deliverable after publication acceptance].

## 3 Results

*3.1 Countries and institutions of affiliation of researchers and topics investigated*

Twenty-nine percent of the authors were affiliated with institutions from the United States, followed by 27% from the United Kingdom, 15% Australia and Netherlands, 14% Germany, and 12% China. Figure 1 sheds light on which countries and which universities are investigating which topics. Universities such as *Beijing Normal* conducts research mostly on *risk governance*.

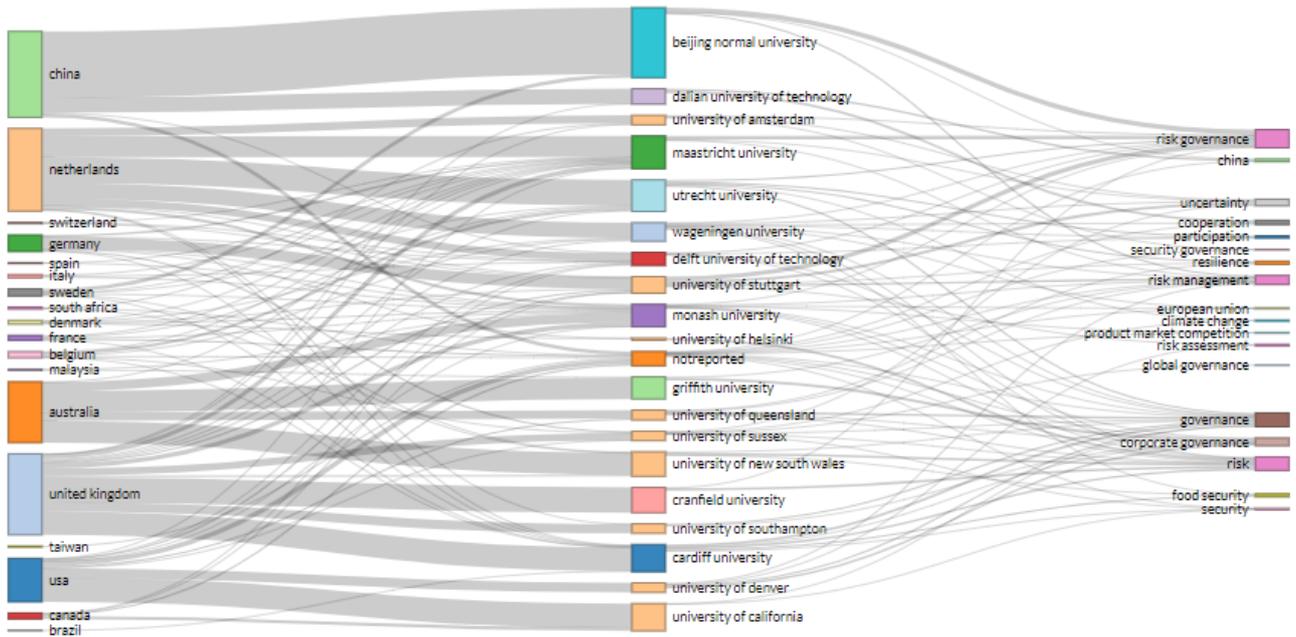

**Figure 1. Countries and institutions of affiliation of researchers and topics investigated 1998-2018. Source: the authors' based on Scopus (2018) and processed with R (R Core Team, 2014) and bibliometrix (Aria & Cuccurullo, 2017).**

*3.2     Superposition map*

Each circle presented in Figure 2 represents a period and the number of different keywords extracted from the articles. The keywords increased from 69 during 1998-2002 to 2,043 during 2013-2018. The number above the horizontal arrows represents the keywords shared between periods. The similarity index between periods (see the number between parentheses) shows that the terms have a low semantic relationship. Finally, the diagonal arrows describe the number of new descriptors in each period. There has been a sustained increase, going from 276 to 1793. It reflects a wide range of relationships with other topics, therefore, interdisciplinary in the field.

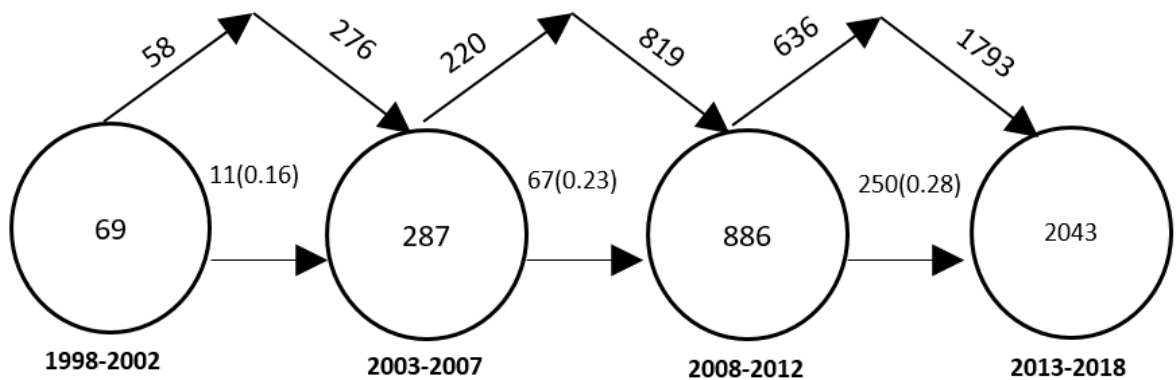

**Figure 2 Key-words superposition. Source: the authors' based on Scopus (2018) and processed with SciMAT (Cobo et al., 2012).**

*3.3     Identification of the central axes of development of the field*

The size of the node is proportional to the number of articles identified by a given keyword. The keywords: *governance'* and *'corporate governance'* stand out during 1998-2002. The emergence of *'governance'* as a central category is reinforced in the following periods. The category *'risk management'* strongly emerged during 2003-2007.

The thematic development of *'corporate governance'* was aimed at integrating aspects related to competition and risk management and risk-taking with agency theory approaches. During 2008-2012 there was the most diversified emergence of new categories with more than thirty analytical categories, with few numbers of articles although suggesting a dispersion of the field. For the same period, new analytical categories associated with the development of governance emerge in public and private settings exemplified by categories such as '*control*' and '*security*' and the leadership of the '*state.*' The last period describes the development of multiple dimensions of governance study, such as '*climate change,*' '*energy,*' '*water,*' '*food security,*' or '*disaster management.*' That confirms the observation above on the higher interdisciplinary knowledge in the field.

In sum, the development of thematic links throughout the periods highlights the establishment of weak links between the different categories where *governance* and *corporate governance* establish the most robust links. The evolution of the conceptual map showed the consolidation of governance and corporate governance in the first half of the period. Also, the diversification of analytical approaches in aspects associated with environmental issues and disaster risk management.

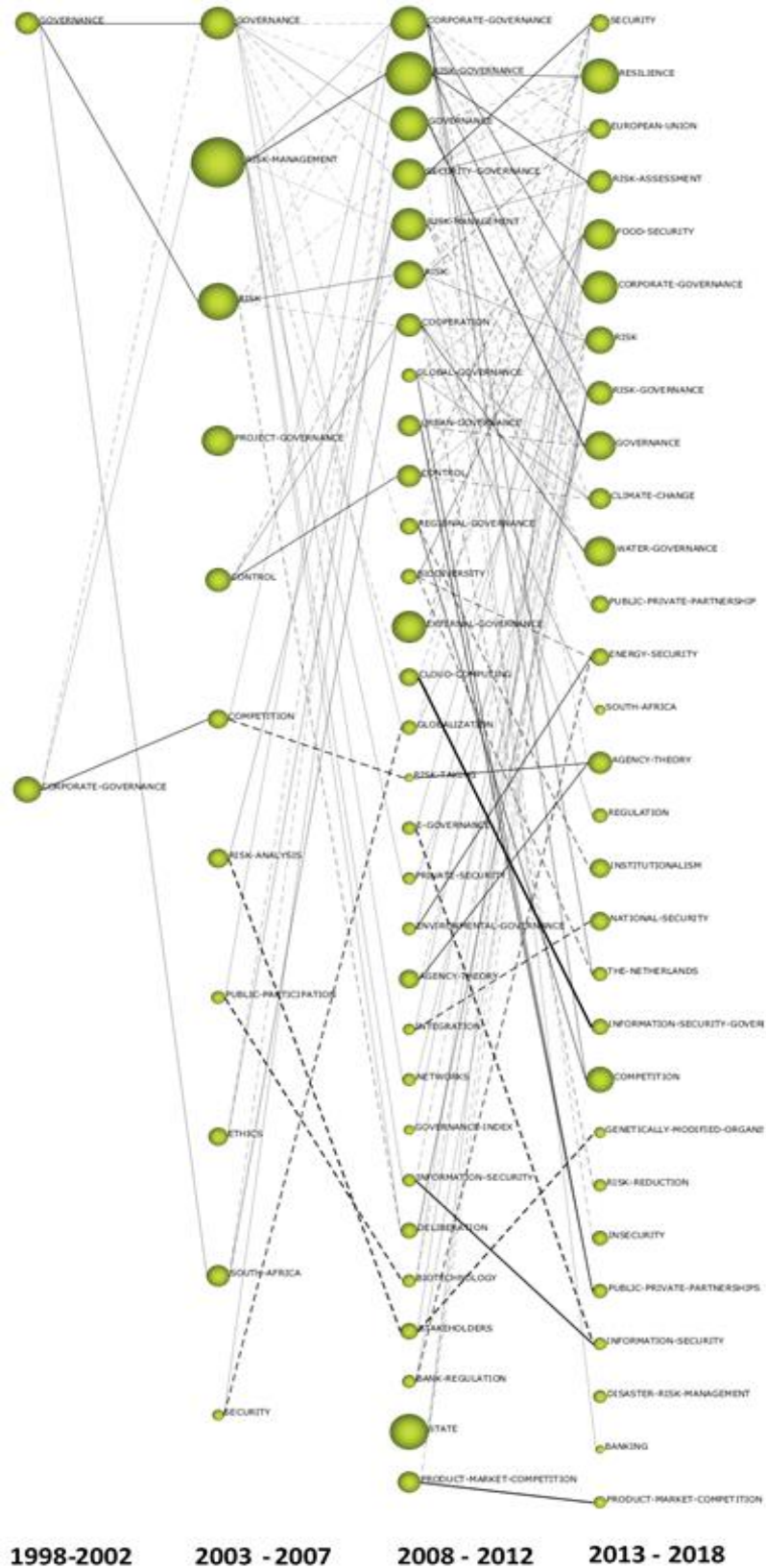

**Figure 3** Thematic evolution of the field 1998-2002, 2003-2007, 2008-2012, and 2018. Source: the authors based on Scopus (2018) and processed with SciMAT (Cobo et al., 2012).

*3.4    Bibliographic coupling network BCN*

Figure 4 presents how the BCN is conducted for an example of two documents. This process was implemented for the total sample of articles. BCN computes the shared references between documents (Zupic & Cater, 2015). Clusters of knowledge emerge when documents share a considerable amount of references or common knowledge. Another advantage of BCN is that both seminal and influential but also new literature, are visible in the network (Zupic & Čater, 2015).

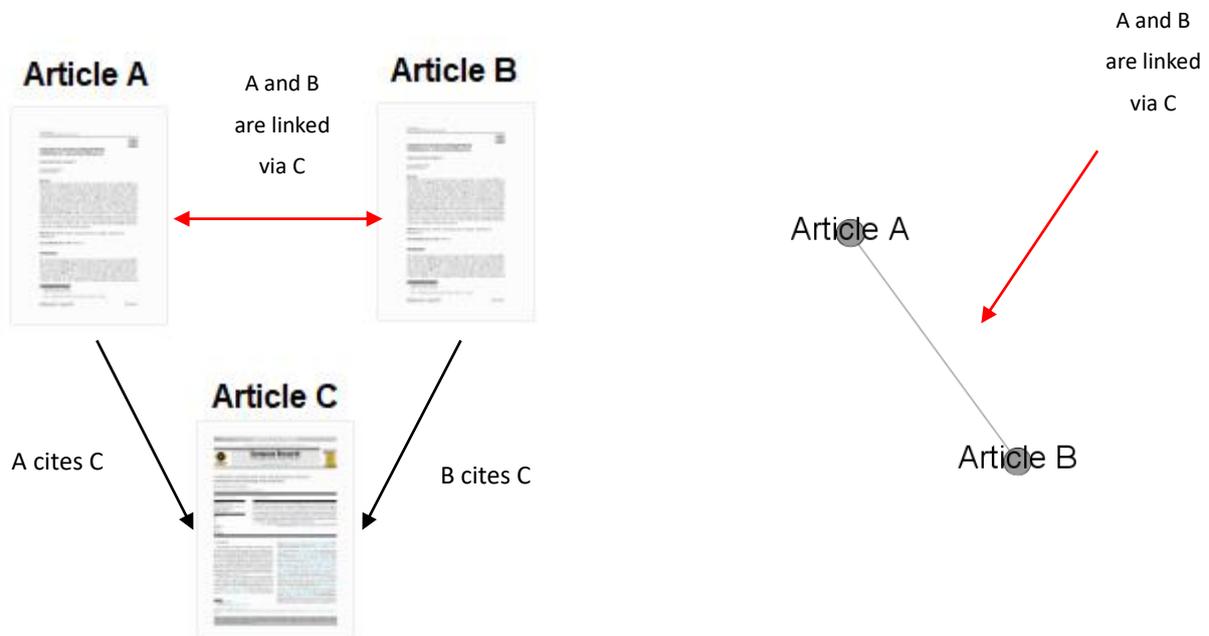

**Figure 4 A conversion example of bibliographic coupling. Two documents are linked if they cite similar references (left) then a network can be modeled after that (right). Source: the authors.**

Figure 5 shows the BCN. The network presents the top-five most representative clusters and 50% of the nodes with the highest betweenness, for improving data visualization. Five-hundred and twenty-nine (529) articles and 92,308 links compose the network, resulting in a density of 0.6. Cluster 3 (pink) is the biggest with 26% of the articles, followed by Cluster 1 (green) with 23%, Cluster 2 (blue) with 21%, Cluster 4 (black) with 14%, and Cluster 5 (orange) with 11%. Nodes' size is proportional to their betweenness. Each node has an ID number that can be used to search it in the dataset available via the QR code (Figure 5). See *supplementary material part 2* for each cluster's top-five articles betweenness and ID number.

The article with the highest betweenness was authored by Elshandidy and Neri (2015) (ID: 438). It studied the influence and impact of corporate governance on risk disclosure practices and market liquidity in the UK and Italy. We proposed clusters' topics based on the top-five betweenness articles. Cluster 3 (pink) deals with the effect of risk governance on bank holding companies adopting risk-appetite practices, decentralized strategies for collaborative crisis communication management between government and public authorities, and policy instruments facing flood impacts on agricultural production in regions such as North-America and Europe. Cluster 1 (green) groups research on social and environmental science, focusing on risk governance for flood management, mainly in Europe. Cluster 2 (blue), similarly to Cluster 3 (pink), is integrated by research on corporate governance, the relevance of non-executive outside directors, and its relationship with risk disclosure practices and market liquidity, particularly in Europe and Africa. Cluster 4 (black) introduces a novel collection of topics related to information and communication technology (ICT) and information security and service quality in, also, a novel set of Asian developing countries such as Malaysia and India. Finally, Cluster 5 (orange) approaches governance for urban planning and infrastructure risks, regarding climate change adaptation policies and the importance of science-based inputs for its design, and the construction of large infrastructure projects such as dams.

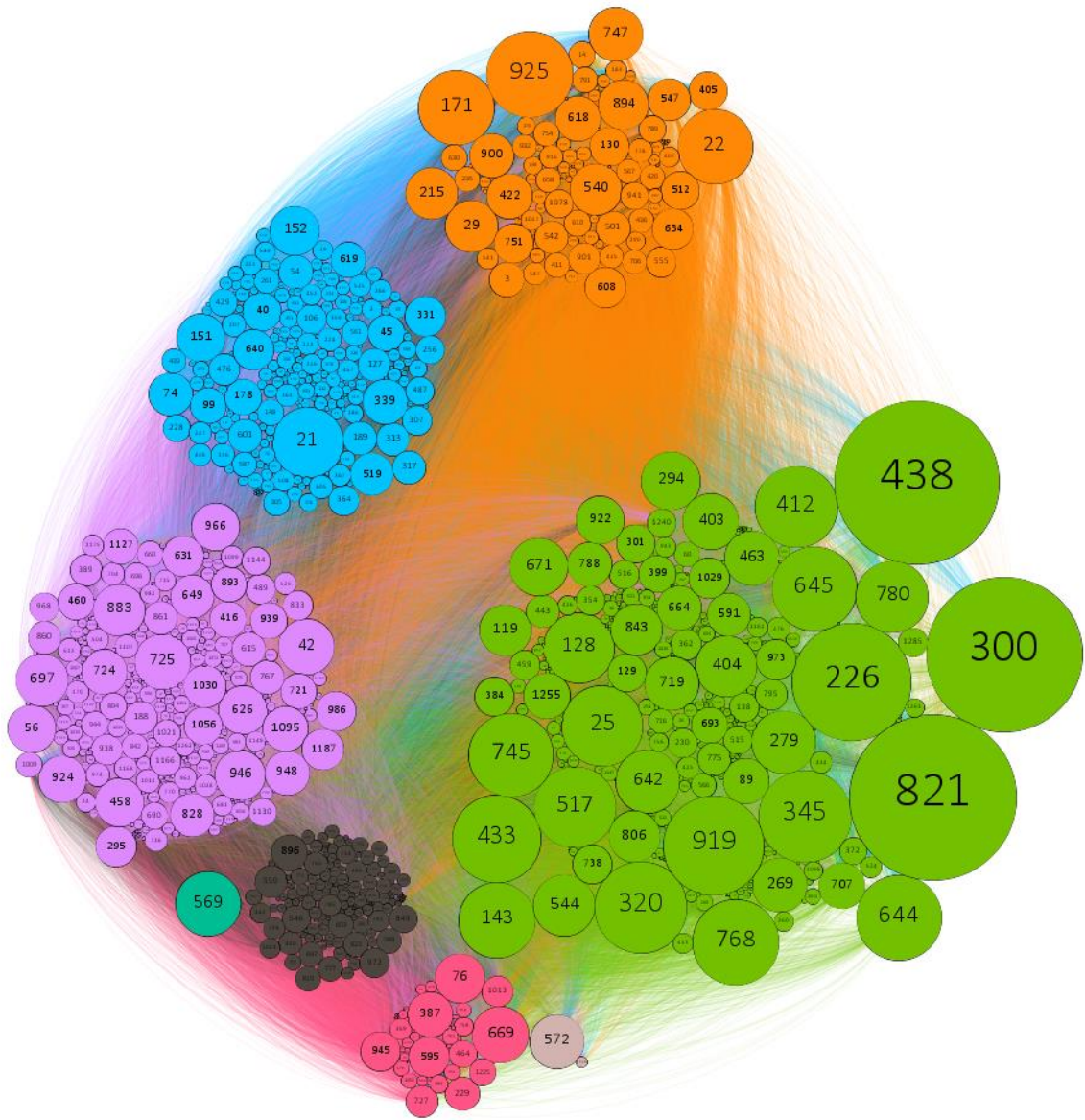

**Figure 5 Bibliographic coupling network. Source: Scopus (2018) processed with R (R Core Team, 2014), bibliometrix (Aria & Cuccurullo, 2017) and Gephi (Bastian et al., 2009).**

In sum, the geopolitical focus has been the global north (Europe and North America). The exception was studies related to ICT and information security and quality in developing Asia. Cluster 1 and 5 showed a common ground related to risk governance for environmental-related issues such as floods and climate change adaptation and the remarkable role of science-based policies. Cluster 2 and 3 are linked by the angle of corporate governance within the financial sector market and regulatory institutions (i.e., norms and rules). Since India handles more than half of the world's IT outsourcing (Frayer & Pathak, 2020) it is clear the highlight of ICT related studies with information security and quality in Cluster 4.

# 4 Discussion

The analyses of keywords superposition, thematic evolution map, and BCN describe the way in which the integrated framework of governance and security, risk, competition, and cooperation has evolved in the last 20 years. The first two analyses showed that corporate governance and risk governance had been consolidated as structural fields. Besides, environmental and technology categories emerged in the last periods examined. These developments describe the conceptual and empirical advancement of governance in dimensions that go beyond public or business management. The research fronts identified by the BCN, adds to the previous findings of the overwhelming focus in problems located in the global north and the relevance of urban planning and the vertical role of science-based policies. We also outlined three paths of further discussion related to the conceptualization, implementation, and level of goals of governance.

## 4.1 Conceptualization dimension

Conceptualization highlights how the empirical base of new knowledge examined enriched the association of governance with other knowledge spheres and actors, i.e., '*polysemic*' according to Baron (2003) and Koechlin (2014). Here, public actors (e.g., the government) interact with the market and civil society through specific objectives. Such *'common-good'* objectives reduce transaction costs, not to say founding democratic agreements for decision-making related to public goods (e.g., water governance). Public and private management have been mediated by institutions (norms and rules) and sustainable development. Ethics and trust, articulate the relationship between stakeholders, and IT is central for the transmission of ideas and information between them. Problems and issues examined range from the most traditional (e.g., public management, business management or public or private finance), to emerging (e.g., climate change, terrorism, food security,), and further (e.g., nanotechnology, biotechnology, biodiversity or urban planning and land use). Figure 6 summarizes this first path of inquiry.

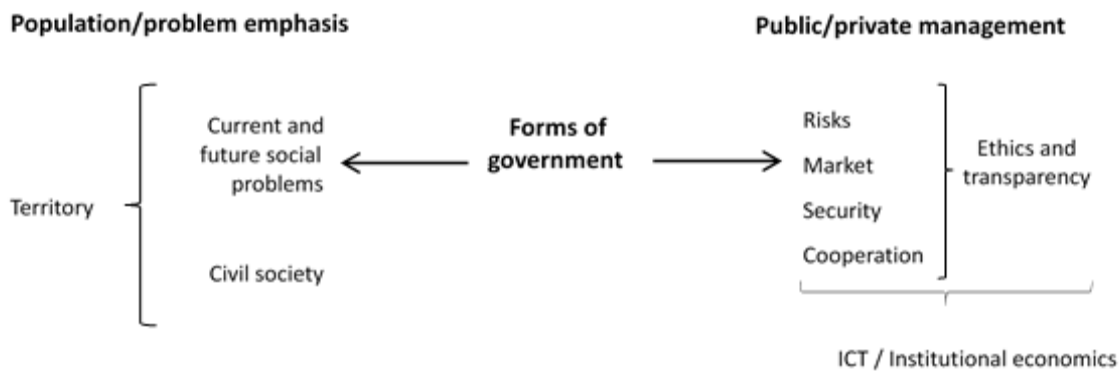

**Figure 6 Conceptual development in the field of governance, risk, security and cooperation. Source: authors' elaboration**

## 4.2 Implementation dimension

Implementation is characterized by diversification in the topics, target populations, and transdisciplinarity. Regarding the first two, results show a greater emphasis on agency actions in the long term that guarantee the well-being of the population (e.g., biotechnology or nanotechnology). These studies are characterized by mitigating environmental risks over time through cooperation between agents involved. Regarding transdisciplinarity, results highlight the contribution of fields such as ICT and institutional economics in the development of control and transparency mechanisms on issues associated with governance, risk, security, and cooperation.

For instance, ICT research is oriented towards the development of channels of access to information in real-time that contributes to the dissemination of information, transparency, and reducing risks. Regarding institutional economics, research is characterized for focusing on reducing transaction costs, managing and controlling externalities, and establishing mechanisms for the dissemination of signals to the market among parties involved through approaches such as agency theory, property rights, and signaling theory (Tang, Lai & Cheng, 2012; Mokhtar & Mellett, 2013; Bae, Masud & Kim, 2018).

*4.3  Level of goals of governance*

Finally, the levels of goals of governance identified were: meta, macro, meso, and micro. Meta-governance refers to the socio-cultural practices that shape the relationship between agents. Also, it contributes to the creation of relationships of trust between institutions and civil society. Macro governance refers to the articulation between institutions and agents that interact according to the forms of public-private government. At the meso-governance level, mechanisms for interaction and control are established to legitimize governance. Finally, micro-governance are the practices of individuals who act as dual subjects, as parts of the form of governments, and who are governed by the articulating principles of the remaining agents and institutions.

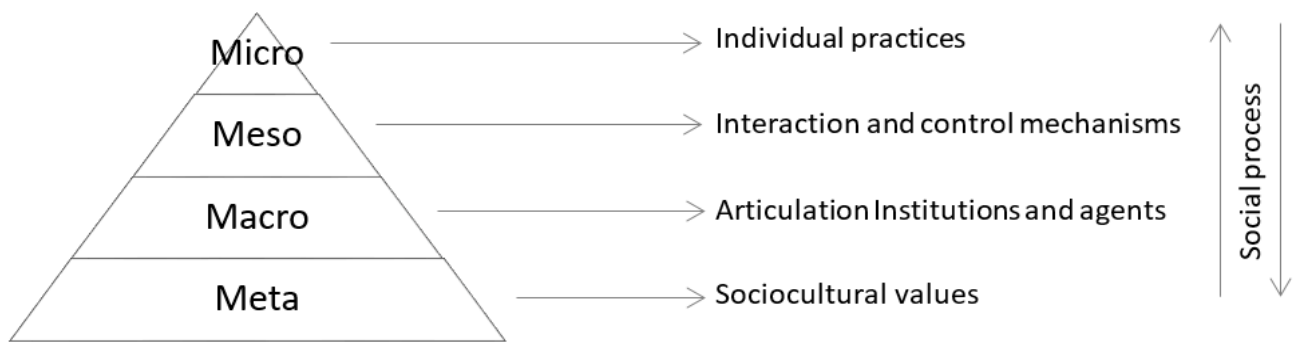

**Figure 7 Level of goals of governance. Source: the authors. Source: authors' elaboration**

# 5   Conclusion

The knowledge mapping analyzed and discussed showed that in the last 20 years, research topics on governance and risk are central, beyond the remaining topics: security, competition, or cooperation. It is essential to notice the emerging role of governance on environmental, urban planning and technology issues, further the bi-nomial public-private sector tensions. The three dimensions here proposed and described (i.e., conceptualization, implementation, and goal levels), enriched the understanding of the interaction mediated by norms/rules between the public sector, the market and the civil society via shared objectives, the long term aims of agency actions for population well-being, and how the goals of governance could vary in reach and scope.

The results presented in this study could serve as a compass for researchers, independent organizations, and decision-making agencies in the public sector, to identify the structure of the evidence and research available for tackling the interdisciplinary and complex problems related to governance, risk, security, competition, and cooperation. Further research could enrich these perspectives trough non-redundant bibliometric technics such as co-authorship analysis for identifying key actors and their connection in the global sphere of knowledge collaboration, examine further variables such as funders and their predilection for specific topics, or the knowledge circulation dynamics in digital media and platforms for social networking using non-traditional metrics (i.e., altmetrics).

**Supplementary material**

**Part 1**

*Network analysis*

The network properties calculated for both co-word and BC networks were: mean degree, density, degree, closeness, betweenness, and eigencentrality, due to their general use in network analysis (Iacobucci et al., 2017). Equation 1 shows how to calculate the mean degree of a network.

$$md = \sum d \ / \ n \ (1)$$

Source: Opsahl, Agneessens, and Skvoretz (2010).

Where *n* is the total of nodes (i.e., points or agents) in a network and *d* is the degree (i.e., number of links a node has). *The mean degree is the average number of links (edges) per node in a network*. Equation 2 shows how to calculate a network's density.

$$den = 2L/n(n-1) \ (2)$$

Source: Scott (1988).

Where *L* is the number of links and *n* the number of nodes. *The density is the proportion of links in a network relative to the total number of links possible: a density of 1 means the whole network is connected*. Equation 3 shows how to calculate a node's degree.

$$C_D(p_k) = \sum_{i=1}^{n} a(p_i, p_k) \ (3)$$

Source: Opsahl, Agneessens, and Skvoretz (2010).

Where *n* is the number of nodes and *a(p_i, p_k)*=1 if and only if the node *i* and *k* are linked; *a(p_i, p_k)*=0 otherwise. *The higher the degree, the more connected the node*. Equation 4 shows how to calculate a node's closeness.

$$C_c(p_k) = \sum_{i=1}^{n} d(p_i, p_k)^{-1} \ (4)$$

Source: Opsahl, Agneessens, and Skvoretz (2010).

Where *d(p_i, p_k)* is the shorter path between nodes *p_i* y *p_k*.. *The closeness indicates how close a node is to all other nodes in the network*. *The higher the closeness, the closer a node is to all other nodes*. Equation 5 shows how to calculate a node's betweenness.

$$C_B(p_k) = \sum_{i<j}^{n} \frac{g_{ij}(p_k)}{g_{ij}} ; i \neq j \neq k \ (5)$$

Source: Opsahl, Agneessens, and Skvoretz (2010).

Where *g_{ij}* is the shorter path that links nodes *p_i* and *g_{ij}(p_k)* is the shorter path that links nodes *p_i* and *p_j p_k*. *Betweenness centrality measures how important a node is to the shortest paths through the network. The higher value, the higher a node's betweenness.* Equation 6 shows how to calculate a node's eigencentrality.

$$Ax = \lambda x, \quad \lambda x_i \sum_{j=1}^{n} a_{ij} x_j, \quad i = 1, \dots, n \ (6)$$

Source: Ruhnau (2000).

This equation describes the eigenvector centrality *x* in two equivalent ways, as a matrix equation and a sum. The eigenvector centrality of a node is proportional to the sum of the eigenvector centrality of the nodes to which that node is linked to. Λ is the vector value of *A* (adjacency matrix) and *n* the number of nodes. *An eigenvector value of 1 corresponds to the node with the highest quality of links in the network regardless of its degree*.

**Part 2**

| | | | Mean degree | 348.9 | | | |
|---|---|---|---|---|---|---|---|
| | | | Density | 0.6 | | | |
| Cluster | ID | Article author, year, and journal | | Degree | Closeness | Betweenness | Eigencentrality |
| 1 | 42 | petrova mh. 2018. rev int stud. | | 826 | 0.793 | 0.003 | 0.992 |
| 1 | 725 | best j. 2013. third world q. | | 788 | 0.773 | 0.003 | 0.966 |
| 1 | 697 | krieger k. 2013. regul governance. | | 766 | 0.759 | 0.003 | 0.953 |
| 1 | 946 | assmuth t. 2010. sci total environ. | | 667 | 0.719 | 0.003 | 0.849 |
| 1 | 56 | kaufmann m. 2018. j flood risk manage. | | 670 | 0.718 | 0.003 | 0.854 |
| 2 | 438 | elshandidy t. 2015. corp gov: int rev. 2015 | | 733 | 0.749 | 0.010 | 0.885 |
| 2 | 821 | waweru nm. 2012. corp ownersh control. | | 786 | 0.772 | 0.009 | 0.916 |
| 2 | 300 | stein v. 2016. j bus econ. | | 722 | 0.746 | 0.009 | 0.869 |
| 2 | 226 | buchwald a. 2017. manage decis econ. | | 647 | 0.706 | 0.007 | 0.767 |
| 2 | 919 | apostolou ak. 2011. corp ownersh control. | | 506 | 0.651 | 0.006 | 0.601 |
| 3 | 21 | gontarek w. 2018. financ mark inst instrum. | | 747 | 0.755 | 0.004 | 0.935 |
| 3 | 152 | georg l. 2017. j manage gov. | | 673 | 0.720 | 0.003 | 0.859 |
| 3 | 151 | larsson ol. 2017. risk. hazards crisis public policy. | | 797 | 0.775 | 0.003 | 0.993 |
| 3 | 74 | hurlbert m. 2018. int j river basin manage. | | 784 | 0.772 | 0.002 | 0.969 |
| 3 | 519 | danov m. 2014. j priv int law. | | 811 | 0.784 | 0.002 | 0.993 |
| 4 | 896 | aziz ka. 2011. int j manage pract. | | 598 | 0.680 | 0.002 | 0.812 |
| 4 | 550 | bahl s. 2014. inf manage comput secur. | | 561 | 0.667 | 0.002 | 0.786 |
| 4 | 972 | sigurjonsson to. 2010. corp gov. | | 499 | 0.647 | 0.002 | 0.693 |
| 4 | 546 | hã£â¤berli c. 2014. mod law rev. | | 755 | 0.758 | 0.002 | 0.965 |
| 4 | 849 | hofferberth m. 2011. bus polit. | | 768 | 0.759 | 0.001 | 0.974 |
| 5 | 925 | corfee-morlot j. 2011. clim change. | | 789 | 0.770 | 0.004 | 0.962 |
| 5 | 22 | ansell c. 2018. risk. hazards crisis public policy. | | 668 | 0.711 | 0.003 | 0.859 |
| 5 | 747 | renn o. 2013. sustainability. | | 672 | 0.719 | 0.003 | 0.848 |
| 5 | 29 | escuder-bueno i. 2018. j risk res. | | 783 | 0.767 | 0.003 | 0.973 |
| 5 | 894 | van asselt mba. 2011. j risk res. | | 602 | 0.685 | 0.003 | 0.781 |

Source: the authors based on Scopus (2018) and processed with R (R Core Team, 2014), bibliometrix (Aria & Cuccurullo, 2017) and Gephi (Bastian et al., 2009).